\documentclass[a4paper,USenglish,cleveref, autoref, pdfa]{lipics-v2021}

\title{Formalizing Higher-Order Termination in Coq}

\titlerunning{Formalizing Higher-Order Termination in Coq}

\author{Deivid Vale}
{Institute for Computation and Information Sciences, Radboud University, The Netherlands \and \url{https://www.cs.ru.nl/~deividvale/}}
{deividvale@cs.ru.nl}
{https://orcid.org/0000-0003-1350-3478}
{Author supported by NWO project “ICHOR”, NWO 612.001.803/7571.}

\author{Niels van der Weide}
{Institute for Computation and Information Sciences, Radboud University, The Netherlands \and \url{https://nmvdw.github.io}}
{nweide@cs.ru.nl}
{https://orcid.org/0000-0003-1146-4161}
{}

\authorrunning{D. Vale and N. van der Weide}

\Copyright{Deivid Vale and Niels van der Weide}

\ccsdesc[500]{Theory of computation~Logic and verification}
\ccsdesc[300]{Theory of computation~Equational logic and rewriting}

\keywords{higher-order rewriting, Coq, termination, formalization}

\nolinenumbers %
\hideLIPIcs
\EventEditors{John Q. Open and Joan R. Access}
\EventNoEds{2}
\EventLongTitle{42nd Conference on Very Important Topics (CVIT 2016)}
\EventShortTitle{CVIT 2016}
\EventAcronym{CVIT}
\EventYear{2016}
\EventDate{December 24--27, 2016}
\EventLocation{Little Whinging, United Kingdom}
\EventLogo{}
\SeriesVolume{42}
\ArticleNo{23}

\usepackage{amsmath,amssymb,amsfonts,amscd,amsbsy,mathtools}

\usepackage{bussproofs}

\usepackage{xcolor}
\usepackage{xcolor-material}
\usepackage{stmaryrd}
\usepackage{dashbox}

\theoremstyle{definition}
\newtheorem{defi}{Definition}[section]
\newtheorem{exa}[defi]{Example}

\theoremstyle{plain}
\newtheorem{thrm}[defi]{Theorem}

\newcommand{\bye}[1]{}
\newcommand{\etal}{\emph{et~al.}}

\newcommand\isafor{\textsf{Isa\kern-0.15exF\kern-0.15exo\kern-0.15exR}}
\newcommand\ceta{\textsf{C\kern-0.15exe\kern-0.45exT\kern-0.45exA}}

\newcommand{\sigFont}[1]{\mathsf{#1}} %
\newcommand{\varFont}[1]{#1} %
\newcommand{\sTp}[1]{\mathsf{#1}} %
\newcommand{\funcFont}[1]{\mathtt{#1}}

\newcommand{\signature}{\mathcal{F}}            %
\newcommand{\var}{\mathbb{X}}                  %

\newcommand{\sortset}{\mathcal{B}}              %

\newcommand{\simpTypes}{\mathcal{ST}_\sortset}
\newcommand{\arrType}{{\Rightarrow}}

\newcommand{\rules}{\mathcal{R}}

\newcommand{\wm}[1]{\mathrel{\mathcal{WM}_{#1}}}

\newcommand{\trs}{{(\mathsf{ar},\rules)}}

\newcommand{\termTp}[2]{\mathsf{Tm}(#1, #2)}

\newcommand{\interpret}[1]{\llbracket #1\rrbracket}

\newcommand{\arrz}{\to_\rules}

\newcommand{\ar}{\funcFont{ar}}

\newcommand{\listvar}{q}

\newcommand{\abs}[2]{\lambda #1.#2}

\newcommand{\aType}{A}
\newcommand{\bType}{B}
\newcommand{\cType}{C}

\newcommand{\aFunc}{\sigFont{f}}

\newcommand{\aVar}{\varFont x}

\newcommand{\aTerm}{s}
\newcommand{\bTerm}{t}

\newcommand{\aCtx}{\Gamma}

\newcommand{\nat}{\sTp{nat}}
\newcommand{\lst}{\sTp{list}}

\newcommand{\nil}{\sigFont{nil}}

\newcommand{\cons}{\sigFont{cons}}

\newcommand{\map}{\sigFont{map}}

\input{coq.sty}
\begin{document}

\maketitle

\begin{abstract}
    We describe a formalization of higher-order rewriting theory
    and formally prove that an AFS is strongly normalizing if %
    it can be interpreted in a well-founded domain.
    To do so, we use Coq, which is a proof assistant based on dependent type theory.
    Using this formalization,
    one can implement several termination techniques,
    like the interpretation method or dependency pairs,
    and prove their correctness.
    Those implementations can then be extracted to OCaml,
    which results in a verified termination checker.
\end{abstract}

\section{Introduction}
Termination, while crucial for software correctness, is difficult to check in practice.
For this reason, various tools have been developed that can automatically check termination of a given program.
Furthermore, such tools are often based on first-order rewriting, for example,
\textsf{AProVE} and \textsf{NaTT}.
However, whereas termination techniques evolved and became more sophisticated over the years,
termination checkers have become more complicated as a result.
Since the proofs outputted by these tools tend to be large,
it is difficult for humans to check for correctness.

For termination tools based on higher-order rewriting \cite{kop:20},
this is even more the case.
Higher-order rewriting is an extension of first-order rewriting
in which function symbols might also have arbitrary functions as argument.
Such an extension gives extra expressiveness,
because higher-order rewriting allows one to deal with higher-order functional programs.
It also lies at the backend of various theorem provers,
it can be used for code transformation in compilers, and more.
However, with greater expressiveness comes a more difficult theory
and more elaborated tools that check for termination are needed.
This leads to the following problem:
how can we formally guarantee the correctness of the answers given by such tools?

This problem already got quite some attention.
One approach is not to prove that the termination checker is correct,
but instead, to check the correctness of the certificates it outputs.
Such an approach was taken by Contejean \etal~\cite{evelyne:et-al:07} and in CoLoR,
developed by Blanqui, Koprowski, and others \cite{BlanquiK11}.
Those tools take as input a certificate produced by a termination checker,
and then the proof assistant Coq checks whether it can reconstruct a termination proof.
If Coq says yes, then the result was actually correct.
Note that Contejean \etal~deal with the first order case,
while CoLoR also includes some methods applicable to higher-order rewriting.

In Isabelle \cite{wenzel2008isabelle},
the library \isafor~contains numerous results about first order rewriting.
Such results can be used to automatically verify the termination of functions defined in Isabelle using a tool that produces certificates of termination \cite{krauss:et-al:11}.
In addition, algorithmic methods were implemented using this library, and from that,
they extracted the verified termination checker called \ceta~\cite{rene:ster:09}.
More recent efforts have also been put in formalizing other tools,
for instance, Thiemann and Sternagel developed a verified tool for certifying AProVE’s termination proofs of LLVM IR programs \cite{rene:ster:09,haslbeck:thiemann:21}.

Our goal is to develop a verified termination checker for higher-order rewrite systems.
To do so, we start by formalizing basic theory on rewriting in the proof assistant Coq \cite{barras:et-al:97}.
After that,
we can implement several algorithms,
such as the interpretation method, dependency pairs, path orders,
and use the theory to prove their correctness.
With all of that in place,
we can use extraction to obtain an OCaml implementation that satisfies the given specifications \cite{letouzey2008extraction}.
Note that the extraction mechanism of Coq was proven to be correct using MetaCoq~\cite{sozeau2020metacoq}.

In this paper, we discuss a formalization of the basic theory of higher-order rewriting.
To do so, we start by discussing signatures for algebraic functional systems in Section \ref{sec:hor}.
The definitions introduced there are the basic data types of the tool.
In Section \ref{sec:inter}, we discuss the main theorem that guarantees the correctness of semantical methods.
We conclude with an overview of what we plan to do in Section \ref{sec:future}.

\subparagraph*{Formalization.}
All definitions and theorems in this paper have been formalized with the Coq proof assistant \cite{barras:et-al:97}.
The formalization is available at \url{https://github.com/nmvdw/Nijn}.
Links to relevant definitions in the code are highlighted as \dashbox{dashed boxes}.

\section{Higher-Order Rewriting}\label{sec:hor}
In this work, we consider \textit{Algebraic Functional Systems} (AFSs), a slightly simplified form of a higher-order functional language introduced by Jouannaud and Okada~\cite{jou:oka:91}.
This choice gives an easy presentation as it combines algebraic definitions in a first-order style with a functional mechanism using $\lambda$-abstractions and term applications.
It is also the higher-order format used in the \href{http://termination-portal.org/wiki/Termination_Competition}{\dashbox{Termination Competition}}.
This gives us access to a variety of higher-order systems defined in the \textit{termination problems database}~\cite{tpdb}, which we plan to use as benchmark for our tool in the future.

In this section, we define the notion of AFS, and we describe how we formalized this notion in Coq.
The definitions given here correspond to those usually given in the literature~\cite{fuhs:kop:12}
while the definitions in Coq deviate slightly.
Note that since we want to extract our formalization to an OCaml program, we use a \emph{deep embedding}, which is similar to the approach taken in \ceta \ and CoLoR \cite{BlanquiK11,rene:ster:09}.
This means that for all relevant notions, such as terms and algebraic functional systems, we define types that represent them.
On the contrast, one could also use a \emph{shallow embedding} where the relevant operations are represented as functions \cite{evelyne:et-al:07}.

Before we can say what an AFS consists of, we need to define types, well-typed terms, and rewrite rules.
We start by looking at types.

\begin{defi}[\href{https://nmvdw.github.io/nijn/nijn.Syntax.Signature.Types.html\#ty}{\dashbox{Types}}]\label{def:stypes}
    Given a set $\sortset$ of base types.
    The set $\simpTypes$ of \textbf{simple types} is inductively built from $\sortset$ using
    the right-associative type constructor $\arrType$. Formally,
    \begin{lstlisting}
Inductive ty (B : Type) : Type :=
| Base : B -> ty B
| Fun : ty B -> ty B -> ty B.
\end{lstlisting}
\end{defi}

In the formalization, we use \lstinline{A1 $\longrightarrow$ A2} to denote function types.
The next step is the notion of well-typed term, which are terms together with a typing derivation.
This is where the \textit{pen-and-paper} presentation deviates from the formalization.
While we present named terms, the formalization, however, makes use of \textit{De Bruijn indices}, so variables are \textit{nameless}.
\bye{
    We present named terms, that is, we assume a set $\var$ of variables in which we hold type information in a context.
    The formalization, however, makes use of De \textit{Bruijn indices}, so terms are \textit{nameless}.
}
This choice affects how to formalize variable environments and terms.
Notwithstanding, the two presentations are equivalent. The former gives a more pleasant informal presentation
and the latter makes the formalization easier.
Note that in the formalization, we use the terminology ``context'' instead of variable environment: that is because in type theory, variable environments usually are called contexts.

\begin{defi}[\href{https://nmvdw.github.io/nijn/nijn.Syntax.Signature.Contexts.html\#con}{\dashbox{Var Env.}}]\label{def:ctx}
    A \textbf{variable environment} $\aCtx$ is a finite list of variable type declarations of the form $\aVar : \aType$ where the variables $x$ are pairwise distinct.
    Formally,
    \begin{lstlisting}
Inductive con (B : Type) : Type :=
| Empty : con B
| Extend : ty B -> con B -> con B.
\end{lstlisting}
\end{defi}

In what follows we sugar \lstinline{Extend A C} as \lstinline{A ,, C} to denote the extension of variable environments.
Now we can define the type of variables as the positions in an environment.

\begin{lstlisting}
Inductive var {B : Type} : con B -> ty B -> Type :=
| Vz : forall (C : con B) (A : ty B),
	var (A ,, C) A
| Vs : forall (C : con B) (A1 A2 : ty B),
	var C A2 -> var (A1 ,, C) A2.
\end{lstlisting}

The set of terms is generated by a set of \emph{function symbols} and the usual constructors from the simply typed lambda calculus (abstraction, application, and variables).
Since terms come together with a typing derivation, we must also know the type of each function symbol.
Hence, when we define the notion of well-typed term, we must assume that we have a function $\ar$ that assigns to each function symbol a type.

\begin{defi}[\href{https://nmvdw.github.io/nijn/nijn.Syntax.Signature.Terms.html\#tm}{\dashbox{Terms}}]\label{def:typed-terms}
    Suppose, we have a map $\ar : \signature \rightarrow \simpTypes$.
    For each type $\aType$,
    we define the set $\termTp{\aCtx}{\aType}$ of \textbf{well-typed terms} of type $\aType$
    in the variable environment $\aCtx$
    by the following clauses:
    \begin{itemize}
        \item given a function symbol $\aFunc \in \signature$,
              we have a term $\aFunc \in \termTp{\aCtx}{\ar(\aFunc)}$;
        \item for each variable $\aVar : \aType \in \aCtx$
              we have a term $\aVar \in \termTp{\aCtx}{\aType}$;
        \item given a term $\aTerm \in \termTp{\aCtx \cup \{\aVar : \bType\} }{\cType}$,
              we get a term $\abs{\aVar}{\aTerm} \in \termTp{\aCtx}{\bType \arrType \cType }$;
        \item given a term $\aTerm \in \termTp{\aCtx}{\bType \arrType \cType}$ and $\bTerm \in \termTp{\aCtx}{\bType}$,
              we have $\aTerm\ \bTerm \in \termTp{\aCtx}{\cType}$.
    \end{itemize}
    Given a variable environment $\aCtx$ and a type $\aType$, we write $\aCtx \vdash \aTerm : \aType$ to denote that $\aTerm$ is a term of type $\aType$ in $\aCtx$.
\end{defi}

To formalize the notion of terms, we make use of dependent types,
which belong to the core features of proof assistants based on Martin-L\"of Type Theory.
This is because $\termTp{\aCtx}{\aType}$ does not only depend on sets,
but also on their inhabitants.
It is formally expressed as below:
\begin{lstlisting}[mathescape=true]
Inductive tm {B : Type} {F : Type} (ar : F -> ty B) (C : con B) : ty B -> Type :=
| BaseTm : forall (f : F),
    tm ar C (ar f)
| TmVar : forall (A : ty B),
    var C A -> tm ar C A
| Lam : forall (A1 A2 : ty B),
    tm ar (A1 ,, C) A2 -> tm ar C (A1 $\longrightarrow$ A2)
| App : forall (A1 A2 : ty B),
    tm ar C (A1 $\longrightarrow$ A2) -> tm ar C A1 -> tm ar C A2.
\end{lstlisting}

Substitutions play an important role in the formalism,
we use them to instantiate rewrite rules and the rewriting relation.

\begin{defi}[\href{https://nmvdw.github.io/nijn/nijn.Syntax.Signature.TermSubstitutions.html\#sub}{\dashbox{Substitution}}]\label{def:subst}
    A \textbf{substitution} $\gamma$ is a finite
    type-preserving map from variables to terms.
    The application of $\gamma$ to $\aTerm$ is denoted by $\aTerm \gamma$.
\end{defi}

\begin{defi}[\href{https://nmvdw.github.io/nijn/nijn.Syntax.Signature.RewritingSystem.html\#rew}{\dashbox{Rewrite rule}}]
    A \textbf{rewriting rule} is a pair of terms $\ell \to r$ of the same type.
    Given a set of rewriting rules $\rules$,
    the \textbf{rewrite relation} induced by $\rules$ on the set of terms is the smallest monotonic relation
    that is stable under substitution and contains both all elements of $\rules$ and $\beta$-reduction.
    That is, it is inductively generated by:
    \[
        \begin{array}{rcllcrcll}
            \ell\gamma      & \arrz & r\gamma    & \text{if}\ \ell \to r \in \rules &  &
            u\ s            & \arrz & u\ t       & \text{if} \ s \arrz t                 \\
            s\ u            & \arrz & t\ u       & \text{if}\ s \arrz t             &  &
            \abs{x}{s}      & \arrz & \abs{x}{t} & \text{if}\ s \arrz t                  \\
            (\abs{x}{s})\ t & \arrz & s[x:=t]    &                                  &  &
                            &       &            &                                       \\
        \end{array}
    \]
\end{defi}
Putting all this data together, gives us the \textit{formalized} notion of \emph{signature for an AFS}.
\begin{defi}\label{def:signature_afs}
    [\href{https://nmvdw.github.io/nijn/nijn.Syntax.Signature.html\#afs}{\dashbox{Signature}}]
    A \textbf{signature for an AFS} consists of the following ingredients:
    \begin{itemize}
        \item a set $\sortset$ of base types and a set $\signature$ of function symbols;
        \item a map $\ar : \signature \to \simpTypes$
              and a set $\rules$ of rewriting rules.
    \end{itemize}
\end{defi}

\begin{remark*}[Nomenclature]
    It is worth mentioning that Definition \ref{def:signature_afs} deviates from the standard notion of signature as is commonly defined in the rewriting community.
    There, a signature (it may be typed or not) is usually defined as a set of symbols used to generate terms.
    Our choice to do so differently is mainly motivated by formalization purposes.
    Since our formalization only uses objects that come from a signature as in Definition \ref{def:signature_afs},
    this notion plays the same generating role as that of the standard notion in the literature.

    In the formalization, one needs to provide the ingredients from Definition \ref{def:signature_afs} to construct an AFS.
    On pen-and-paper, we denote such an object by $(\sortset, \signature, \ar, \rules)$,
    and if the set of base types is clear from the context, we denote an AFS just as $\trs$.
    This is a kind of compatibility between the formalization and the pen-and-paper presentation.
\end{remark*}

Now let us look at a simple example.
The base types are lists and natural numbers.
The function we look at applies a function $F$ to every element of a list $q$.
Note that this example relies on the fact that we have higher-order types.
\begin{exa}\label{exa:map}
    We build lists using the constructors $\vdash \nil : \nat$ and $\vdash \cons : \nat \arrType \lst \arrType \lst$.
    The rules for $\map$ can be typed using the variable environment $F : \nat \arrType \nat, \aVar : \nat$.
    \begin{align*}
        \map(F, \nil) & \to \nil & \map(F, \cons(\aVar, \listvar)) & \to \cons(F \aVar, \map(F, \listvar))
    \end{align*}
\end{exa}

\section{Higher-Order Interpretation Method}
\label{sec:inter}

The goal of our tool is to check for termination,
and thus theorems that give conditions for strong normalization
form the core of the formalization.

Within the higher-order framework,
several methods have been developed,
and our focus is on the so-called \emph{semantical methods}:
to prove an AFS is terminating we need to find a well-founded interpretation domain
such that $\interpret{\aTerm} > \interpret{\bTerm}$, whenever $\aTerm \arrz \bTerm$.
This is achieved by orienting each rule in $\rules$, that is,
$\interpret{\ell} > \interpret{r}$ for all rules $\ell \to r$ in $\rules$.
The idea was first introduced by {van de Pol}~\cite{pol:96}
(in the context of HRS) as an extension of the first-order semantic interpretation method.
Later, these semantic methods got extended to AFSs by {Fuhs and Kop}~\cite{fuhs:kop:12} with
a special focus on implementation, and it is part of Wanda,
a termination tool developed by Kop~\cite{kop:20}.

In this section, we discuss the main definitions and theorems for such methods.
Our notion of well-founded interpretation domain is that of an \emph{extended well-founded set}.
These are sets together with two ordering relations:
a well-founded strict $>$ and a quasi-order $\ge$ compatible with it.
More precisely, compatible orders are defined as follows.

\begin{defi}[\href{https://nmvdw.github.io/nijn/nijn.Prelude.CompatibleRelation.html\#CompatRel}{\dashbox{Compatible Order}}]\label{def:compatible-order}
    An \textbf{extended well-founded set} is a tuple $(X, >, \ge)$ consisting of a set $X$,
    a well-founded relation $>$ on $X$, and
    a transitive and reflexive relation $\ge$ on $X$,
    such that, for all $x, y, z \in X$, the following conditions hold:
    \begin{itemize}
        \item
              $x > y$ implies $x \ge y$;
        \item
              if $x > y$ and $y \ge z$, then we have $x > z$;
        \item
              if $x \ge y$ and $y > z$, then we have $x > z$.
    \end{itemize}
\end{defi}

To interpret base types, we just need to give an extended well-founded set for each base type.
However, we also need to be able to interpret function types.
We use \emph{weakly monotonic functions} for that.
These are functions that preserve the quasi-ordering $\geq$.

\begin{defi}
    Given extended well-founded sets $X$ and $Y$,
    a \textbf{weakly monotonic function} $f$ is a function
    $f : X \to Y$ such that for all $x, y \in X$, if $x \ge y$, then we have $f(x) \ge f(y)$.
    If we also have $f(x) > f(y)$ whenever we have $x,y \in X$ such that $x > y$, then we say $f$ a \textbf{strongly monotonic function}.
\end{defi}

Given two extended well-founded sets $X$ and $Y$,
we can construct another well-founded set whose elements are weakly monotonic functions.
Using those, we can define the interpretation $\wm{\aType}$ of a simple type $\aType$
the same way as {Fuhs and Kop}~\cite{fuhs:kop:12}.
For a full interpretation, we also should interpret terms.
The necessary data for that is more complicated, and for the details we refer the reader to~\cite{fuhs:kop:12}.

Briefly said, a
\emph{\href{https://nmvdw.github.io/nijn/nijn.Interpretation.WeaklyMonotonicAlgebra.html}{weakly monotonic algebra}} is needed to interpret types and symbols in $\signature$.
However, to be able to use it as a reduction ordering additional information is required.
Namely, we need a symbol $@^A$, for each type $A$,
satisfying a strictness condition
to represent term-application;
and symbols in $\signature$ has to be interpreted as strongly monotonic functionals.
From such an extended algebra, we get maps $\interpret{\cdot}$
that send terms of type $\aType$ to elements of $\wm{\aType}$.
Using these notions, we can formally prove the main theorem.

\begin{thrm}[\href{https://nmvdw.github.io/nijn/nijn.Interpretation.WeaklyMonotonicAlgebra.html\#afs_is_SN_from_Alg}{\dashbox{Compatibility Theorem}}]
    Let an AFS $\trs$ and an extended weakly monotonic algebra for it be given.
    If for each rewriting rule we have $\interpret{\ell} > \interpret{r}$,
    then the AFS is strongly normalizing.
\end{thrm}

\section{Future Work}
\label{sec:future}
We briefly discussed the basics of our formalization of higher-order rewriting in Coq.
Up to now, we formalized the basic data structures, namely types, terms, and signatures, and
give a formal proof for the compatibility theorem.
This theorem is necessary to guarantee the correctness of algorithms that
use semantic methods to check for strong normalization.

The next step is to formalize rule removal and actual instances algorithms.
We plan to start with higher-order polynomial interpretation.
After a full formalization of the tool-chain,
we can extract an OCaml program from the Coq implementation,
which is guaranteed to satisfy the proven specifications.
As a result, we will get a fully verified tool that checks for termination of higher-order programs.
\newpage

\bibliography{main}

\end{document}